\documentclass[twocolumn,showpacs]{revtex4}
\usepackage{graphicx,subfigure}
\usepackage{tabularx}
\begin{document}
\title{Fast realization of a spatially correlated percolation model}
\author{Hongting Yang\footnote{email:yht@whut.edu.cn}}
\affiliation{School of Science, Wuhan University of Technology, Wuhan 430070, P.R. China}
\author{Stephan Haas}
\affiliation{Department of Physics and Astronomy, University of Southern California, Los Angeles, CA 90089-0484}
\date{\today}

\begin{abstract}
We propose two schemes to achieve fast realizations of spatially correlated percolation models.
The schemes are shown to be efficient in complementary regimes of correlation phase space.
They are combined with a generalized Newman-Ziff algorithm to numerically determine the percolation
thresholds of two-dimensional lattices in the presence of correlations. It is found that the spatial
correlations affect only a relatively small part of phase space.
\end{abstract}

 \maketitle

\section{introduction}
Much of the existing literature on percolation has focused on the uncorrelated case~\cite{stauffer}, where each lattice
site or bond is independently occupied with probability $p$ or empty with probability $1-p$. Some of the occupied
sites/bonds form clusters, and with more sites or bonds occupied, these clusters grow larger. For finite lattices with
periodic boundary conditions, if a cluster grows large enough to wrap around the lattice, wrapping percolation
occurs~\cite{PM2003}. Analogously, for finite lattices with free boundaries, the appearance of a spanning cluster touching
two opposite boundaries indicates the occurrence of spanning percolation along the direction perpendicular to the two
boundaries.

The correspondence of phase transition with percolation has attracted much interest~\cite{sahimi1994,hunt2009}.
Applications range widely from quarks and gluons to galaxies~\cite{guttner1986,armesto1996,sarkar2010,seiden1990}, and from
epidemics to networks~\cite{davis2008,son2012,lai2010,sergey2010,agliari2011,bizhani2011}. In particular, the notion of
discontinuity in explosive percolation has stimulated a number of interesting recent
studies~\cite{achlio2009,costa2010,radicchi2010,araujo2011,riordan2011}. For the square lattice, it is believed that the
value of the percolation threshold is unique, although its exact value remains unknown for site
percolation~\cite{feng2008,lee2008,yang2012}.

While most of these previous studies do not include spatial correlations, this constraint appears overly restrictive. For
example, phase transitions between solid, liquid and gaseous states take place under different conditions. A theory with
unique percolation threshold can not be applied to describe multiple phase transitions in a unified way. Effective
percolation models without interactions between sites or bonds are incomplete~\cite{sergey2010}. Such interactions, whether
attractive or repulsive, generate spatially correlated site and bond distributions. However, in contrast to uncorrelated
percolation models, it is more complicated to efficiently generate spatially correlated systems~\cite{S2005}.

In this brief report, we examine percolation models with spatial correlations caused by attractions
between occupied sites. We will address two issues. The first question is how to generate spatially
correlated distributions efficiently. In the following section, we present two schemes which work well in complementary parameter regimes.
The second question is how to find the value of a percolation threshold. A generalized Newman-Ziff algorithm provides the tool for
this second task~\cite{NZ2001}. This will be discussed in the third section.

\section{Model and Algorithm}

Here we discuss how to build spatially correlated lattices by considering compactness among occupied sites. It is assumed
that there is a bond between all neighboring sites. In terms of the bond length (the lattice spacing), the least number of
bonds traversed to reach one site from another site is defined as the distance between the two sites. If two sites satisfy
the correlation condition, {\em i.e.} the distance between the two sites is no greater than a preset value, we say that the
two sites are spatially correlated. We use $d$ to represent this preset value, and hence the correlation condition is given
by $d=1, 2, 3, \ldots, L$. Obviously, $d=1$ corresponds to the most compact distribution, and $d=L$ represents a completely
uncorrelated random distribution.

\begin{figure}[h]
\centering \subfigure[\ $d=1$]{\label{f1a}\includegraphics[width=0.2\textwidth]{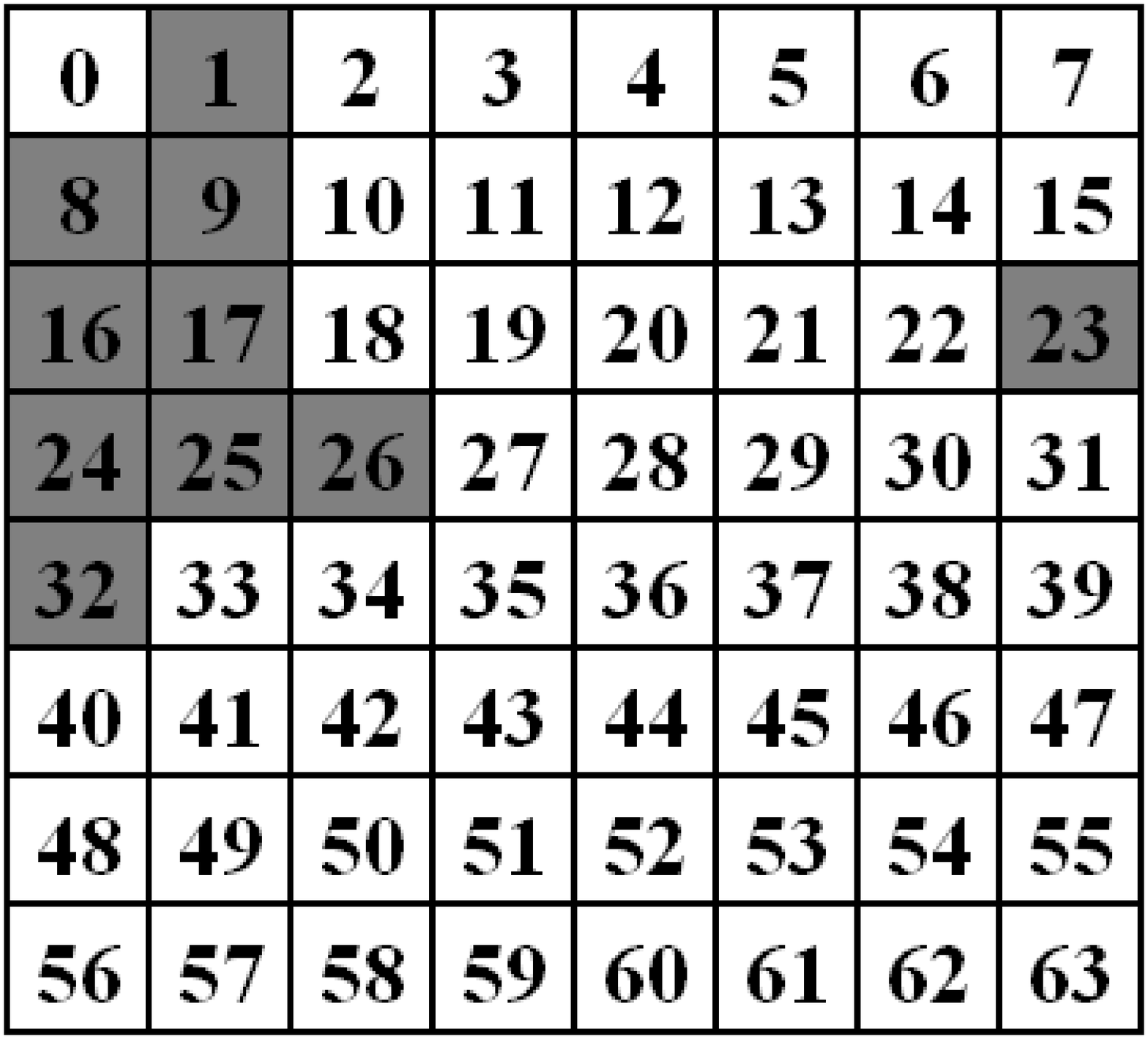}} \subfigure[\
$d=2$]{\label{f1b}\includegraphics[width=0.2\textwidth]{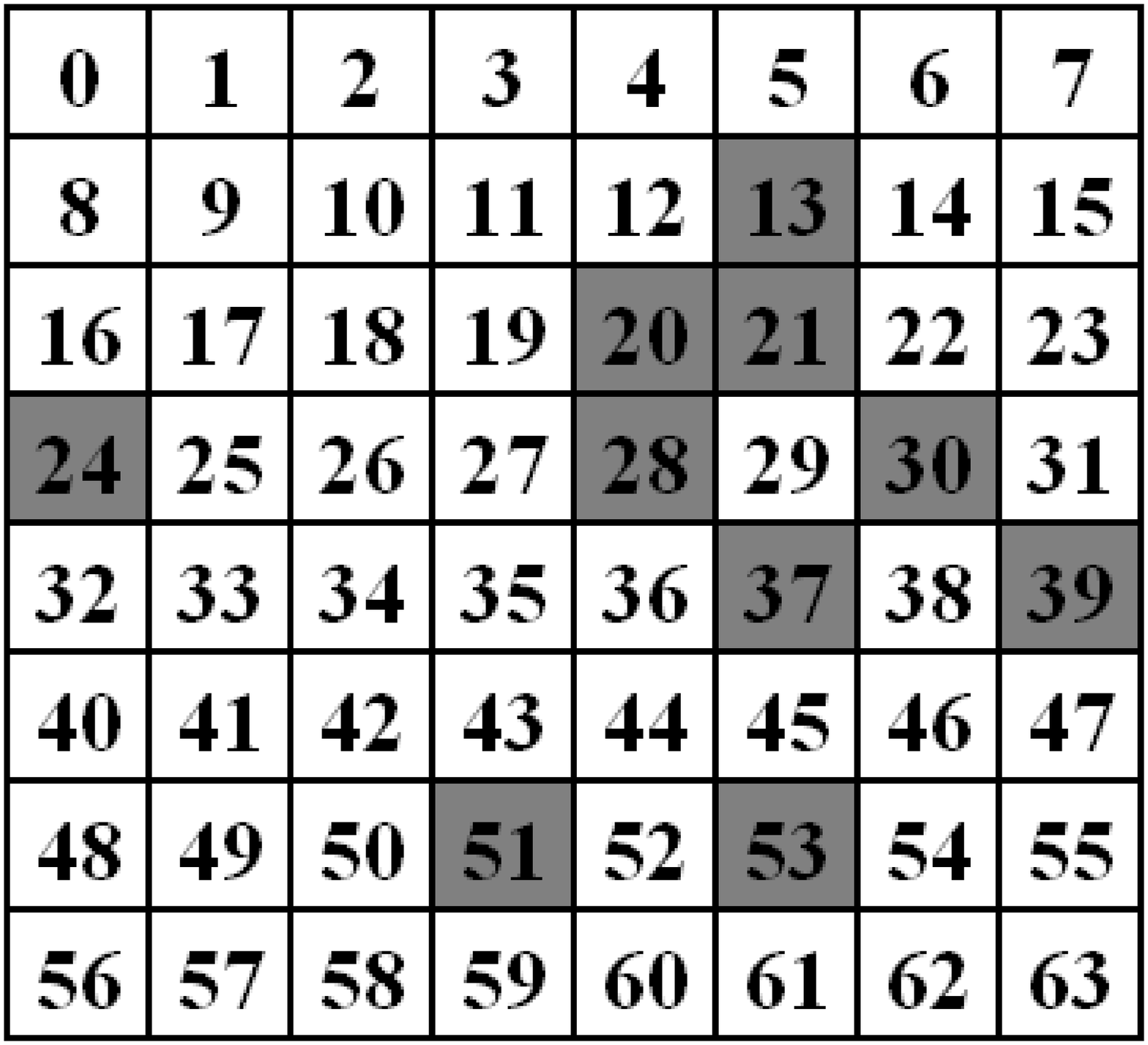}}
\caption{Two examples of spatially correlated distributions on a periodic square lattice with $L=8$.
Ten occupied sites shaded in (a) and (b) are shown for $d=1$ and $d=2$ respectively.\label{f1}}
\end{figure}

Let us now contemplate algorithms how to generate spatially correlated distributions for a square lattice with linear
dimension $L$. Only the first occupied site is chosen completely randomly. The distance between the next site to be
occupied and at least one of the already occupied sites is set to be no greater than $d$. This way, each site occupied
later is spatially correlated to (at least) one of the previously occupied sites. For $d=1$, since this is the most compact
distribution, the effective attraction between the occupied sites is the strongest, and in the process of occupying sites
one by one, there is only one cluster. Two examples of spatially correlated distributions for $d=1$ and $d=2$ on square
lattices with $L=8$ are shown in Fig.~\ref{f1a} and \ref{f1b}.

We now attempt to answer the question of how to generate a spatially correlated percolation model most efficiently.
While this may not be a pressing issue for small lattices as those shown in Fig.~\ref{f1}, for large lattices with $L$ up to 256,
inefficient methods to generate spatially correlated distributions of occupied sites will take exceedingly long running time.
The key to numerical efficiency is when we should test the correlation condition. In general, there are two schemes, which are discussed below.

In scheme A, we test the correlation condition $after$ we randomly choose an empty site, which is chosen to be occupied if
it satisfies the correlation condition. Otherwise, we continue to randomly choose other empty sites, until we find one that
fulfills the correlation condition. Obviously, for a randomly chosen site, greater $d$ implies a greater probability to
satisfy the correlation condition. In contrast, a smaller $d$ means that the empty sites allowed by the correlation
condition are distributed in a smaller area around the occupied sites, and it will therefore take a longer time to find a
correlated empty site among all the empty sites. Scheme A can be summarized as follows.
\begin{description}
  \item[(1)] Randomly choose one empty site.
  \item[(2)] If this site is spatially correlated to one of the previously occupied sites, occupy this site; otherwise, go to step (1).
\end{description}
Repeat steps (1) to (2) until all sites are occupied or wrapping percolation occurs.

In scheme B, the correlation condition is tested $before$ we randomly choose an empty site. We test the correlation condition, and find
all available empty sites spatially correlated to the newly occupied site. We then randomly choose one empty site to occupy directly from
this list of available empty sites. This scheme works effectively for the case of small $d$ on large lattices, whereas for large $d$ on a
large lattice, it will take a longer time to test the correlation condition and find the spatially correlated empty sites. Scheme B can be summarized as follows.
\begin{description}
  \item[(1)] Occupy one site randomly chosen from the list of correlated empty sites, which are spatially correlated to the previously occupied sites.
  \item[(2)] Refresh the list of correlated empty sites by adding extra empty sites spatially correlated to the newly occupied site in step (1).
\end{description}
These steps are repeated until all sites are occupied or wrapping percolation occurs.

To show the great difference between the efficiency of the two schemes in calculation time, an example of a lattice $L=256$
is considered in Table~\ref{t1}. The number of runs of the algorithm for each scheme is taken as $2\times 10^4$. The
computations are implemented on a desktop PC with a CPU clock speed 2.6 GHz and memory 1.96 GB. When $d=1$, scheme B takes
about 336 seconds, while scheme A takes about 260 days. When $d\le 16$, scheme B takes shorter time than scheme A. When
$d\ge32$, the former takes longer time than the latter; when $d=128$, the former takes about 65 hours while the latter
takes only 260 seconds. Obviously, scheme A is not effective when $d$ is small, and scheme B works less effectively when
$d$ is large. For the case of a completely uncorrelated random distribution, $d=L$, scheme B will take about 8 days, while
scheme A takes about 210 seconds, which is close to 170 seconds spent in the Newman-Ziff algorithm. Similar results exist
for other lattices with different $L$. With increasing $L$, the difference of the two schemes in the calculation time
increases too.

In general, in scheme B, the computation time $T_L\sim nL^2$ for fixed $d/L\ge{1\over4}$; for fixed $L$, $T_L$ increases
with increasing $d$, $T_L\sim d^{c_1}$, where $c_1=1.1,1.3,1.4,1.4$ for $L=32,64,128,256$ respectively. In scheme A, the
computation time $T_L\sim nL^2$ for fixed $d$; for fixed $L$, $T_L$ decreases with increasing $d$, $T_L\sim d^{-c_2}$,
where $c_2=1.7,2.0,2.1,2.3$ for $L=32,64,128,256$ respectively. Obviously, from the point of saving computation time, any
scheme alone is not appropriate to the calculation for the whole range of the values of $d$'s. The two schemes should be
combined to give a fast realization of the spatially correlated percolation model.

\begin{table*}[htbp]
  \centering
  \caption{Comparison of calculation time for the two schemes at different $d$'s, on the lattice with $L=256$.
  The number of runs of the algorithm for each scheme is $n=2\times10^4$.}\label{t1}
\begin{tabularx}{\textwidth}{@{\extracolsep{\fill}}llllllllll}
  \hline\hline
         $d$ & 1     & 2     & 4     &      8 &     16 &    32 &   64  & 128    & 256   \\ \hline
    B-scheme & 336 s & 492 s & 885 s & 30 min & 85 min & 4.7 h & 17 h  & 65 h   & 8 d   \\ \hline
    W-scheme & 260 d & 60  d & 11  d & 42 h   & 6.4 h  & 1   h & 700 s & 260 s  & 210 s \\ \hline\hline
\end{tabularx}
\footnotetext{Here, d=day(s), h=hour(s), min=minute(s), s=second(s).}
\end{table*}

\section{Extraction of Percolation Threshold}

On a lattice with $N$ sites, the Newman-Ziff algorithm states that a state with $n+1$ occupied sites is achieved by adding
one extra randomly chosen site to a state with $n$ occupied sites. An observable $Q(p)$ is binomially expanded in terms of
the measured quantities $\{Q_n\}$
\begin{equation}
 Q(p)=\sum_{n=0}^N{N\choose n}p^n(1-p)^{N-n}Q_n.
\end{equation}
Such observables $Q$ can be the probability of cluster wrapping, mean cluster size, correlation length, and so on. Spatial
correlations affect the values of $Q_n$, and thus the value of $Q(p)$. There are four types of probabilities $R_L(p)$ of
cluster wrapping on the periodic square lattice of $N=L\times L$ sites. $R_L^{(e)}$ is the probability of cluster wrapping
along either the horizontal or vertical directions, or both; $R_L^{(1)}$, around one specified axis but not the other axis;
$R_L^{(b)}$, around both the horizontal and vertical directions; $R_L^{(h)}$ and $R_L^{(v)}$, around the horizontal and
vertical directions, respectively. The four wrapping probabilities satisfy the equations
\begin{eqnarray}
  R_L^{(b)} &=& R_L^{(e)}-2R_L^{(1)}, \\
  R_L^{(h)} &=& R_L^{(e)}-R_L^{(1)},
\end{eqnarray}
only two of which need to be measured. In general, given the exact value of $R_\infty(p_c)$, the solution $p$ of the
equation
\begin{equation}
  R_L(p)=R_\infty(p_c),
\end{equation}
gives a very good estimator for the threshold $p_c$ in percolation theory. However, for spatially correlated percolation
models, the exact value of $R_\infty(p_c)$ is unknown in advance, so we can not obtain the value of $p_c$ by solving the
equation. However, we can obtain  $p_c$ from $R_L^{(1)}(p)$ by virtue of its non-monotonicity. We use Machta's
method~\cite{NZ2001,machta96}, instead of the alternative criterion for two-dimensional wrapping
percolation~\cite{yang2012}, to save computation time.

\section{Results}

All computations are implemented on the same desktop PC. We take $d=1, 2, 4, \ldots, L/2, L$, and restrict ourselves to
calculating $p_c$. We choose scheme B when $d\le4$ for $L=32$, $d\le8$ for $L=64$ and $L=128$, $d\le16$ for $L=256$. In
the other cases, we choose scheme A instead. Each time of running the program, the algorithm is implemented for $n$ runs to output
the value of $p_c$, and we run the program $g$ times to estimate its error.

For $L=32$, $d=1$, we take $n=2\times10^8$, running the program once takes about 13 hours. We repeat this ten times,
and obtain the value of percolation threshold and its error: $p_c=0.6026982(32)$. For all other values of $L$ and $d$,
we also take $g=10$, and the number of runs of the algorithm $n$ ranges from $1.4\times10^5$ to $2\times10^8$, and each run takes 8--13 hours.

\begin{figure}[htb]
\centering
\includegraphics[width=0.5\textwidth]{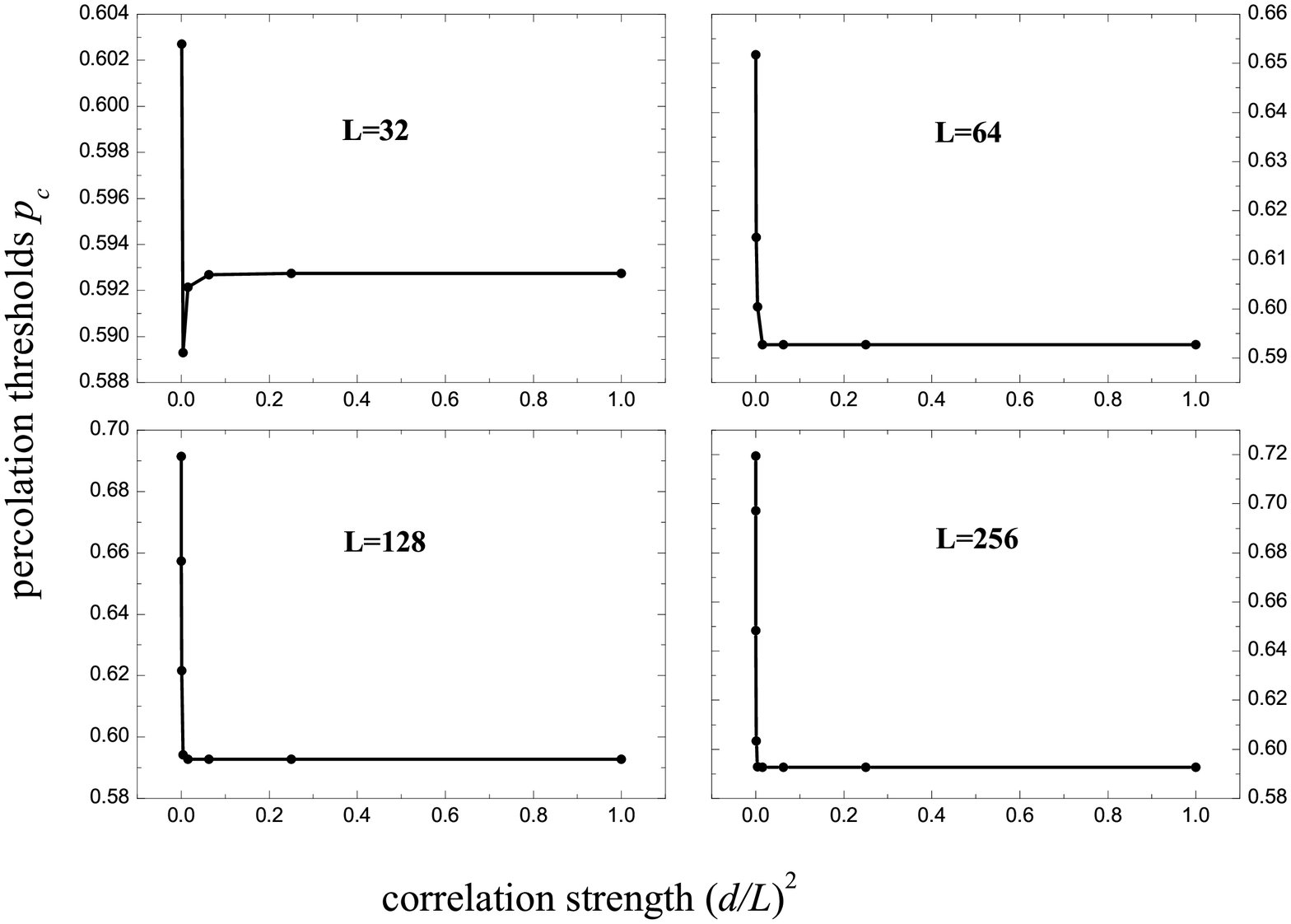}

\caption{\label{f2}Variations of the percolation threshold with the correlation strength for the lattices with $L=32$, 64,
128, and 256.}
\end{figure}

To elucidate the effect of the size of a lattice, the percolation thresholds are shown in Fig.~\ref{f2} as a function of
$(d/L)^2$, instead of $d$. The errors in $p_c$ range from $8.5\times10^{-7}$ to $4.6\times10^{-5}$, which are too small to
be shown. One notices that there is a sharp dip around $d=2$ only in the curve for $L=32$. In the curve for $L=64$,
$p_c=0.5926966(31)$ at $d=8$ is a little bit smaller than the nearby $p_c$ values. With increasing $L$, there are no more signs of dip.
Therefore, the dip in the curve for $L=32$ is believed to be the result of finite size effects.
\begin{figure}[htb]
\centering
\includegraphics[width=0.5\textwidth,height=0.3\textwidth]{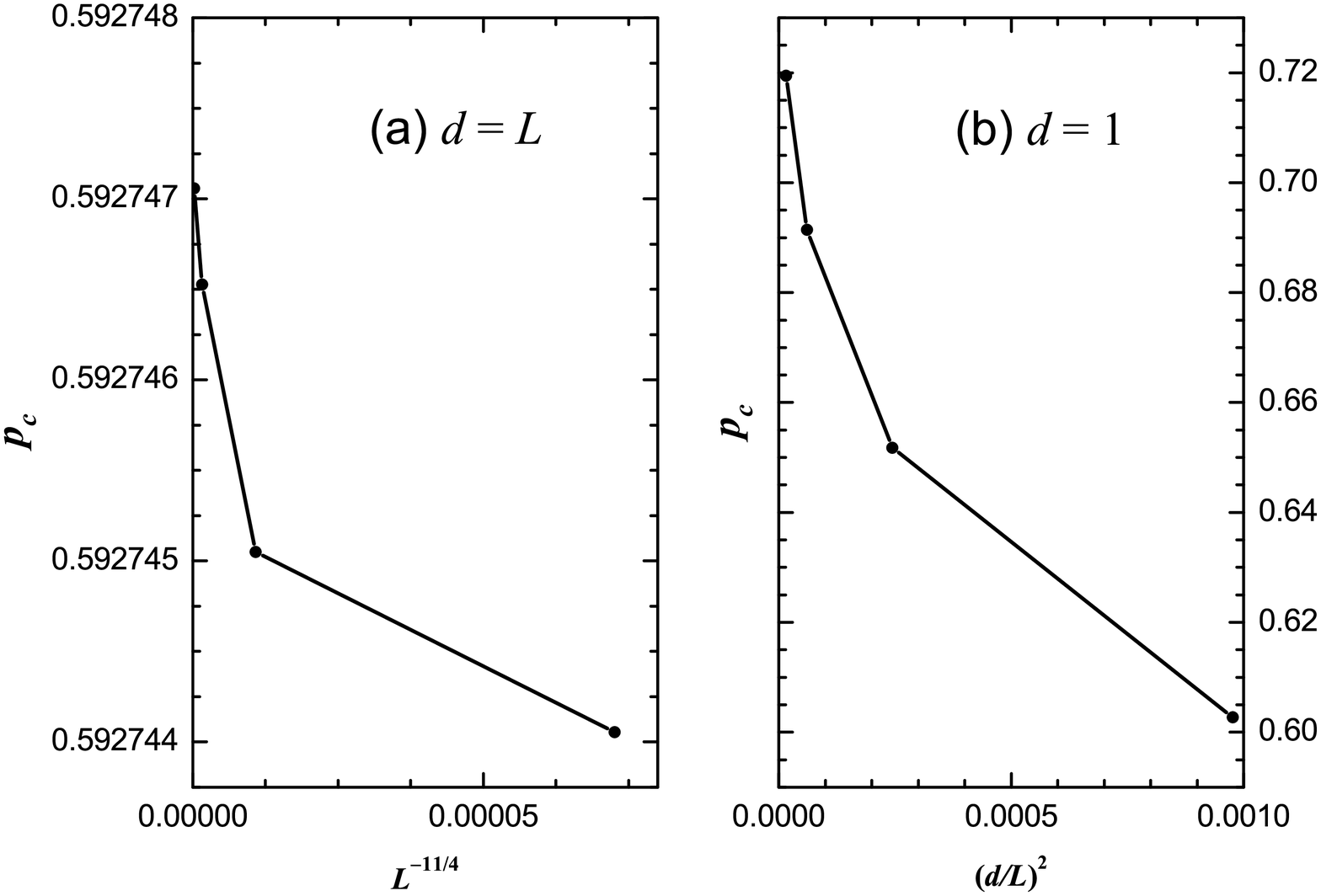}

\caption{The values of percolation thresholds under the extreme conditions, (a) $d=L$ no spatial correlation, and (b) $d=1$
the most compact distribution.}
\end{figure}

The most striking feature in Fig.~\ref{f2} is that the effect of spatial correlation on $p_c$ exists only in a very small regime of
correlation strength $(d/L)^2<0.004$, estimated from the existing data. Except the dip in the curve for $L=32$, the values of $p_c$
nearly keep unchanged in a wide range of correlation strength $0.004<(d/L)^2<1$.

Since $d=L$ corresponds to the case of completely uncorrelated random distribution, the value of percolation threshold converges to
$p_c(\infty)$ according to~\cite{NZ2001}
\begin{equation}
  p_c-p_c(\infty)\sim L^{-11/4}.
\end{equation}
As in Fig.~3(a), the finite size scaling of estimates for $p_c$ gives $p_c(\infty)=0.59274640(52)$, which coincides with
the previous results~\cite{NZ2001,feng2008,lee2008,yang2012}. The strongest correlated case is for $d=1$, i.e. the
strongest effective attraction between occupied sites on an infinite lattice. Linear fitting in Fig.~3(b) gives $p_c(d=1,
L=\infty)=0.701(14)$.

\section{Conclusion}

Scheme A and B are combined to give an efficient realization of spatially correlated lattices, introduced phenomenologically by restricting the distance between sites.
Scheme B is effective especially when there are only a few sites can be chosen to occupy in the process of cluster growing. For any population of sites on a lattice,
one has to choose an appropriate algorithm to achieve the corresponding population.

As for a spatially correlated percolation model, the calculation of critical exponents is of great interest, too. For the purposes of showing the differences of the
two schemes in realizing a site population and the effects of correlation
strength, here we focused on the calculation of percolation thresholds.

Very strong attractions between occupied sites may seriously hinder the occurrence of percolation transition, while
correlations which are not very strong between occupied sites have less effects on the percolation thresholds. This is the
reason why the uncorrelated percolation theory without considering spatial correlations has so many successful
applications.

\end{document}